\documentclass[prd, amsfonts, twocolumn, nofootinbib, showpacs]{revtex4}
\usepackage{graphicx, epsfig}
\usepackage{color}
\usepackage{amsmath}
\usepackage{hyperref}
\usepackage{amsmath}
\usepackage{appendix}
\newcommand{\be}{\begin{equation}}
\newcommand{\ee}{\end{equation}}
\newcommand{\bea}{\begin{eqnarray}}
\newcommand{\eea}{\end{eqnarray}}

\newcommand{\gapp}{\mathrel{\raise.3ex\hbox{$>$}\mkern-14mu \lower0.6ex\hbox{$\sim$}}}
\newcommand{\lapp}{\mathrel{\raise.3ex\hbox{$<$}\mkern-14mu \lower0.6ex\hbox{$\sim$}}}
\def\bbox{{\,\lower0.9pt\vbox{\hrule \hbox{\vrule height 0.2 cm
\hskip 0.2 cm \vrule  height 0.2 cm}\hrule}\,}}

\usepackage{color}
\usepackage{ulem}
\usepackage[usenames,dvipsnames]{xcolor}

\begin{document}
\title{Symmetron Inflation}
\author{ Ruifeng Dong, William H. Kinney and Dejan Stojkovic}
\affiliation{ Department of Physics, University at Buffalo, SUNY, Buffalo, NY 14260-1500}
 %%%%%%%%%%%%%%%%%%%%%%%%%%%%%%%%%%%%%%%%%%%%%%%%%%%%%%%

\begin{abstract}
\widetext
We define a new inflationary scenario in which inflation starts naturally after the Big Bang when the energy density drops below some critical value. As a model, we use recently proposed symmetron field whose effective potential depends on the energy density of the environment. At high densities, right after the Big Bang, the potential for the symmetron is trivial, and the field sits in equilibrium at the bottom of the potential. When the density drops below some critical value, the potential changes its shape into a symmetry breaking potential, and the field starts rolling down. This scenario does not require any special initial conditions for inflation to start. We also construct a concrete model with two fields, i.e. with symmetron as an inflaton and an additional scalar field which describes the matter content in the early universe. For the simplest coupling, the amplitude and shape of the power spectrum are the same as in the single field slow-roll inflation.
\end{abstract}

%%%%%%%%%%%%%%%%%%%%%%%%%%%%%%%%%%%%%%%%%%%%%%%%%%

\maketitle

\section{Introduction}

One of the long standing problems for the inflationary paradigm is the problem of suitable initial conditions required for inflation to start. The ``Extended Copernican Principle", which is an underlying principle in modern physics, implies that we should avoid any fine tuning in our theories, either in choosing the values of parameters in our Lagrangian or values of initial conditions.  From the statistical point of view, the most likely state in a collection of all possible states is the state of maximal entropy. However, the state of maximal entropy usually describes an equilibrium of the system, in which we have ever changing microscopical states, but macroscopically nothing really changes. It is very difficult to imagine any interesting macroscopic dynamics in such a state. To get something going on, we have to take the system far from the equilibrium and let it evolve back to the equilibrium. However, this implies very non-generic initial conditions (corresponding to a state far from the equilibrium) which is at odds with the ``Extended Copernican Principle". In the context of inflation, this problem is acute. In order to start inflation, one has to take the inflaton field far from its minimum and let it roll down the potential. Several approaches have been developed to deal with this fundamental problem \cite{Albrecht:1982wi,Linde:1983gd,Linde:1981mu}; however, it is always useful to look for alternative solutions.

There are two main elements to the initial condition problem for inflation: First, how do we achieve the cosmological homogeneity necessary for inflation to begin \cite{Vachaspati:1998dy}? Second, how do we arrange for the field to be sufficiently displaced from its vacuum state? We focus here on the second question. The aim of this paper is to formulate a model in which inflation starts naturally, without the fine tuning in initial field value. To achieve this, we will study a class of symmetron models, which is just a variant of modified gravity models (scalar-tensor gravity). The scalar symmetron field \cite{Hinterbichler:2010es} has an interesting feature of a density dependent potential. Generically, one can write an effective potential for such a scalar field as
\be \label{Veff}
V_{\rm eff} = \frac{1}{2} \left( \frac{\rho}{M^2} - \mu^2 \right)\phi^2 +   \frac{1}{4} \lambda \phi^4 + const,
\ee
where $\mu$ and $M$ are two mass scales and $\lambda$ is a dimensionless coupling.
In the low density environment where $\rho < \mu^2M^2$ the mass squared term is negative, and the scalar acquires a vacuum expectation value (VEV)
$\phi_0=\frac{\mu}{\sqrt{\lambda}}$. In the high density environment where $\rho > \mu^2M^2$, the effective potential no longer
breaks the symmetry, and the VEV of the scalar field is zero. These models have wide range of applications in late time cosmology, in the context of dark energy and dark matter problem \cite{Hinterbichler:2011ca,Upadhye:2012rc,Brax:2012nk,Winther:2011qb,Pietroni:2005pv}.
We will employ here this interesting mechanism in early universe, in order to formulate the symmetron inflation model.

\begin{figure}[h]
  \centering
\includegraphics[width=4.0in]{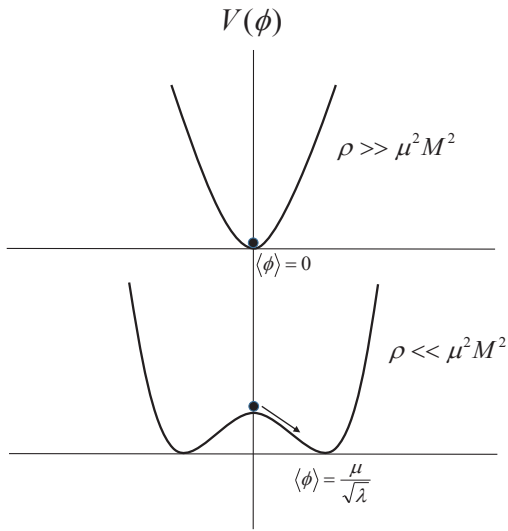}
\caption{This figure shows the evolution of the potential in Eq.~(\ref{Veff}) after the hot Big Bang. Right after the Big Bang, in the high density environment where $\rho \gg \mu^2M^2$, the potential has a trivial shape and the field $\phi$ sits at the bottom. At the critical value of the density $\rho = \rho_c = \mu^2M^2$, the potential changes its shape  and the field $\phi$ starts rolling down (the second order phase transition). When the density drops to $\rho \ll \mu^2M^2$, the potential becomes the standard double well potential.  }
    \label{potential}
\end{figure}

\section{Model}

We will adopt here the following inflationary scenario in which the symmetron field $\phi$ plays the role of inflation. After the Big Bang, the energy densities are very high, i.e. $\rho \gg \mu^2M^2$, the potential has a simple U shape, and the scalar field is sitting at the bottom of the potential. As the universe expands, the density drops, and when it reaches the critical value of $\rho_c = \mu^2M^2$ the potential changes its shape into a characteristic symmetry breaking potential. At that point the field starts rolling down the potential and inflation starts, as shown in Fig.~(\ref{potential}). If the reheating temperature at the bottom of the potential is low enough, the condition $\rho \ll \mu^2M^2$ is never violated. This scenario would not require any unnatural displacement of the inflaton field from its equilibrium.

We will now estimate the values of parameters which would make this scenario plausible.
In order to achieve slow-roll inflation, we need  $\mu << H_{\rm inf}$. On the other hand, at the critical point of transition we have $\frac{1}{3}\mu^2 M^2 = H^2 M_{pl}^2$, which requires $M > M_{pl}$. Here and throughout the paper, $M_{pl}$ stands for the reduced Planck mass. This in turn implies that the the field $\phi$ couples to the matter density $\rho$ very weakly. However, only a mild hierarchy $M \sim 10 M_{pl}$ would be enough to achieve slow roll.

Another requirement we have is $V_{\rm inf} > \mu^4/\lambda$, since the field $\phi$ reaches the value  $\phi_0=\frac{\mu}{\sqrt{\lambda}}$
only at the end of inflation. Therefore
\be V_{\rm inf} > \mu^4/\lambda \sim H^4 M_{pl}^4 / (M^4 \lambda ).
\ee
Since we want $V_{\rm inf} \sim H^2M_{pl}^2$, this condition becomes
\be V_{\rm inf} \sim H^2M_{pl}^2 <  M^4 \lambda.
\ee
Thus the usual requirement $\lambda \ll 1$ can be easily accommodated.

While a simple effective potential (\ref{Veff}) gives the desirable phenomenology, it is also possible to formulate a microscopic theory behind it  \cite{Hinterbichler:2010es}.
We can start from an action
\begin{eqnarray}\label{action}
S=&&\int d^4 x \sqrt{-g} \left(\frac{1}{2}M_{pl}^2 R-\frac{1}{2}g^{\mu \nu}\partial_\mu \phi \partial_\nu \phi -V(\phi)\right)\nonumber\\
&&+\int d^4 x \sqrt{-\tilde{g}} {\cal L}_m \left( \chi , \tilde{g}\right) ,
\end{eqnarray}
with metric signature (-,+,+,+). ${\cal L}_m$ is the Lagrangian for the matter field $\chi$, which couples to the metric $\tilde{g}_{\mu\nu}$ related to the original $g_{\mu\nu}$ by
\begin{equation}\label{gmn}
\tilde{g}_{\mu\nu}=A^2(\phi )g_{\mu\nu}.
\end{equation}
The metrics $g_{\mu\nu}$ and $\tilde{g}_{\mu\nu}$ describe the Einstein and Jordan frames respectively.
Later on in the paper, we will work in the Einstein frame, which is supposed to be FRW. To show the complexity of the coupling between gravity and the two scalar fields, the form of the action in the Jordan frame is given in the Appendix. Note that ${\cal L}_m$ implicitly depends on the field $\phi$ through the metric $\tilde{g}_{\mu\nu}$.
The equation of motion for the field $\phi$ is thus
\be
\Box \phi = V_{,\phi } -A^3 A_{,\phi} \tilde{T},
\ee
where $\tilde{T}$ is the trace of the matter stress-energy tensor in the Jordan frame, with $\tilde{T}_{\mu\nu}=-(2/\sqrt{-\tilde{g}})\delta \left(\sqrt{-\tilde{g}}{\cal L}_m \right)/\delta \tilde{g}^{\mu\nu}$, but the box operator is calculated with the metrics $g_{\mu\nu}$.
In the rest frame of the fluid, the trace of the matter stress-energy tensor depends on the equation of state as $T = (1-3w)\rho $, where $\rho =A^3 \tilde{\rho}$ is independent of $\phi$. This $\rho$ is conserved in Einstein frame and has the usual properties, e.g., redshifts with the scale factor $a$ as $\rho \sim 1/a^{3,4}$ in a matter and radiation dominated universe respectively.

In Einstein frame, for a pressureless source $w=0$, the equation of motion is
\be
\Box \phi = V_{,\phi } + A_{,\phi} \rho.
\ee
Thus, the effective potential is
\be
V_{\rm eff} = V(\phi) + \rho A(\phi).
\ee
 Then the concrete form of (\ref{Veff}) can be obtained with a choice
\begin{eqnarray} \label{A}
V(\phi)&=&\rho_{vacuum}-\frac{1}{2}\mu^2\phi^2+\frac{1}{4}\lambda \phi^4 , \\
A(\phi)&=&1+\frac{1}{2M^2}\phi^2, \nonumber
\end{eqnarray}
where $\rho_{vacuum} = \frac{1}{4}\mu^4/\lambda$ is the vacuum energy density. Since in the simplest form of the action the field $\phi$ does not couple to radiation (the trace $T = (1-3w)\rho $ vanishes for radiation), we need to modify the matter part of the action. To arrange for coupling to radiation, one could for example add an extra factor in front of the Jordan frame Lagrangian
\be \label{pc}
S_m = \int d^4 x \sqrt{-\tilde{g}} B(\phi)  {\cal L}_m \left( \chi , \tilde{g}\right).
\ee
An implicit $\phi$-dependence of the Lagrangian would give the standard coupling to the trace of the stress energy tensor $T$, but the new ingredient now is the $B(\phi)$ prefactor. This factor corresponds to a ``pressure'' coupling. Its contribution to the equation of motion for $\phi$ is proportional to ${\cal L}_m$, which is the pressure. For non-relativistic dust with zero pressure it will not give any additional contribution. One of the concrete realizations consistent with all of the assumed symmetries is
\be
B = 1 + \phi^2/M^2 .
\ee
It is thus possible to achieve the standard slow-roll inflationary scenario that starts in a hot radiation-dominated phase, and then follows the usual inflationary evolution.

\section{Modeling the background for the onset of inflation}

An alternative way to assure the outlined inflationary scenario is to start inflation in a matter dominated epoch. This is not difficult to envision since the energy density of the universe right after the Big Bang could have been dominated by degrees of freedom which are effectively pressureless. Inflation itself will effectively dilute these degrees of freedom and erase any pre-inflationary history of the universe. If the reheating temperature is not high enough to restore the original degrees of freedom, the standard radiation dominated era can be provided after inflation ends at the reheating.

Starting inflation in a matter dominated era will assure coupling between the $\phi$ and local energy density even in the original action in Eq.~(\ref{action}).  We can effectively model the matter dominated epoch with the auxiliary scalar filed $\chi$. As a toy model, we consider a massive scalar field $\chi$, with Lagrangian
\begin{equation} \label{ML}
{\mathcal L}_m = -\frac{1}{2} {\tilde g}^{\mu\nu} \partial_\mu \chi \partial_\nu \chi - \frac{1}{2} m^2 \chi^2,
\end{equation}
which enters the action in Eq.~(\ref{action}).
Coupling between the fields $\phi$ and $\chi$ is provided through the metric ${\tilde g}_{\mu\nu}$ given by Eq.~(\ref{gmn}).
If the field  \(\chi\) is homogeneous, we can ignore field gradients, and the stress-energy for the field will be of the form of a perfect fluid, with
\begin{eqnarray}
&&\rho_\chi = \frac{1}{2} \dot\chi^2 + \frac{1}{2} m^2 \chi^2, \cr
&&p_\chi = \frac{1}{2} \dot\chi^2 - \frac{1}{2} m^2 \chi^2.
\end{eqnarray}
During the matter dominated epoch preceding inflation, the field $\phi$ remains fixed at the origin $\phi = 0$, so that the conformal factor $A$ relating $g_{\mu\nu}$ and ${\tilde g}_{\mu\nu}$ (\ref{A}) is unity, so $\chi$ evolves as a scalar field in a FRW background.
Assuming a flat universe with energy density dominated by the field \(\chi\), we can write the Friedmann Equation as
\begin{equation}
\left(\frac{\dot a}{a}\right)^2 = \frac{1}{3M_{pl}^2} \rho_\chi = \frac{1}{3M_{pl}^2} \left[\frac{1}{2} \dot\chi^2 + \frac{1}{2} m^2 \chi^2\right],
\end{equation}
%where $M_p = M_{\rm Pl} / \sqrt{8 \pi}$ is the reduced Planck mass.
The equation of motion for the field is then
\begin{equation} \label{eomc}
\ddot\chi + 3 \left(\frac{\dot a}{a}\right) \dot\chi + m^2 \chi = 0.
\end{equation}
  If $m^2 \chi \gg 3 H \dot\chi$ the system will be underdamped. Since the pressure is zero, we have $1/2 m^2 \chi^2 = 1/2 {\dot\chi}^2$, so $\dot\chi = m \chi$. Then we see that for $H \ll m$ the system is underdamped
\begin{equation}
\ddot\chi + m^2 \chi \simeq 0,
\end{equation}
with solution
\begin{equation}
\chi = \chi_0 e^{\pm i m t}.
\end{equation}
It is simple to see that this represents a matter-dominated system, since the average energy density and pressure are then
\begin{eqnarray}
&&\left\langle\rho_\chi\right\rangle = \left\langle\frac{1}{2} \dot\chi^2 + \frac{1}{2} m^2 \chi^2\right\rangle = m^2 \chi_0^2 , \cr
&&\left\langle p_\chi\right\rangle = \left\langle\frac{1}{2} \dot\chi^2 - \frac{1}{2} m^2 \chi^2\right\rangle = 0.
\end{eqnarray}
The expansion rate is then
\begin{equation}
\left(\frac{\dot a}{a}\right)^2 = \frac{1}{3M_{pl}^2} \left\langle\rho_\chi\right\rangle = \frac{m^2}{3M_{pl}^2} \chi_0^2,
\end{equation}
so that, from continuity, the amplitude of the field decays adiabatically as
\begin{equation}\label{rc}
\rho_\chi = m^2 \left\langle\chi^2\right\rangle \propto a^{-3}.
\end{equation}
Alternatively, the correct WKB-solution of Eq.~(\ref{eomc}) is $\chi \propto  a^{-3/2} e^{\pm imt}$ (as shown in \cite{AS}) from which
Eq.~(\ref{rc}) follows immediately.

\section{Evolution of the fields before and during inflation}

Combining the action for the symmetron field $\phi$ with the Lagrangian for the field $\chi$ which describes the matter content in the universe, we obtain the complete model.
To verify  that this model has the desired properties discussed above, we numerically calculate the evolution of the coupled equations for $\phi$ and $\chi$ in the dynamical background of an expanding universe. From the action ~(\ref{action}) with the matter Lagrangian ~(\ref{ML}), we can get the Friedmann equation ~(\ref{Friedmann}) and the equations of motion for the two fields, Eqs.~(\ref{eom_phi}) and ~(\ref{eom_chi}).
\begin{equation}\label{Friedmann}
H^2 = \frac{1}{3M_{pl}^2} \left(\frac{1}{2}{\dot{\phi}}^2 + V(\phi) + \frac{1}{2}A^2{\dot{\chi}}^2 + \frac{1}{2} A^4 m^2 {\chi}^2 \right),
\end{equation}
\begin{equation}\label{eom_phi}
\ddot\phi + 3H\dot\phi + V_{,\phi} + A_{,\phi} \left( \frac{1}{2}A{\dot{\chi}}^2 + \frac{1}{2} A^3 m^2 {\chi}^2 \right) = 0,
\end{equation}
\begin{equation}\label{eom_chi}
\ddot\chi + 3H\dot\chi + m^2 A^2 \chi + 2\frac{A_{,\phi}}{A} \dot{\phi}\dot{\chi} = 0.
\end{equation}
In parenthesis on the right-hand side of Eq.~(\ref{Friedmann}), the first two terms correspond to $\rho_\phi$ (i.e. the energy density of $\phi$), while the other two terms correspond to $A(\phi)$ times $\rho_\chi$. As we explained earlier, the conserved energy density for $\chi$ is $\rho_\chi =A^3 \tilde{\rho}_\chi$, where
\be
\tilde{\rho}_\chi = \frac{1}{2} A^{-2}  \dot\chi^2 + \frac{1}{2}  m^2 \chi^2,
\ee
where the factor $A^{-2}$ in front of the kinetic term comes from the metric $\tilde{g}^{\mu \nu}$. On the other hand, the effective density that enters the Friedmann equation (\ref{Friedmann}) is $\bar{\rho}_\chi = A  \rho_\chi = A^4 \tilde{\rho}_\chi$. The reason is that the continuity equation for $\bar{\rho}_\chi$ is modified because of the coupling of $\phi$ to matter fields in Einstein frame:
\be
\dot{\bar{\rho}}_\chi = -3H\bar{\rho}_\chi + \frac{\dot{A}}{A}\bar{\rho}_\chi,
\ee
which integrates to $\bar{\rho}_\chi \sim A(\phi)/a^3$. This is consistent with the fact that in Einstein-frame the mass of non-relativistic matter particles depends on $\phi$ as
$m(\phi) =  m_0 A(\phi)$, where $m_0$ is constant, so the energy density of dust scales as
$\bar{\rho}_\chi = m(\phi)/{\rm volume}  \sim A(\phi)/a^3$.

\begin{figure}[h]
\centering
\includegraphics[width=3.5in]{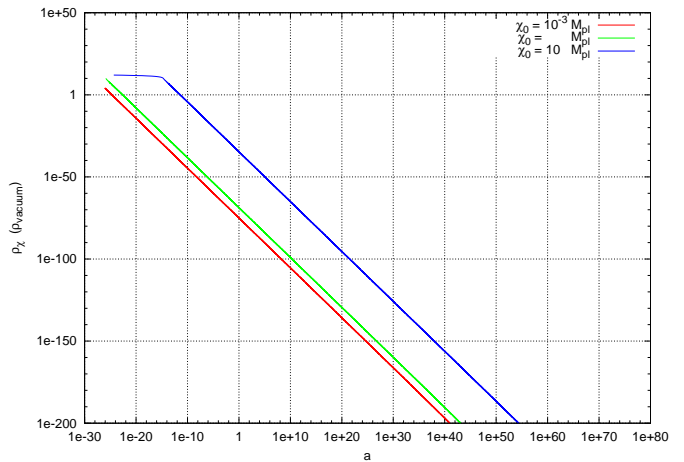}
\caption{The evolution of  $\rho_{\chi}$ for three different initial conditions. After $\chi$ reaches thermal equilibrium, $\rho_\chi$ drops as $a^{-3}$ during the matter dominated era and inflation era.}
\label{rho_chi}
\end{figure}

\begin{figure}[h]
\centering
\includegraphics[width=3.5in]{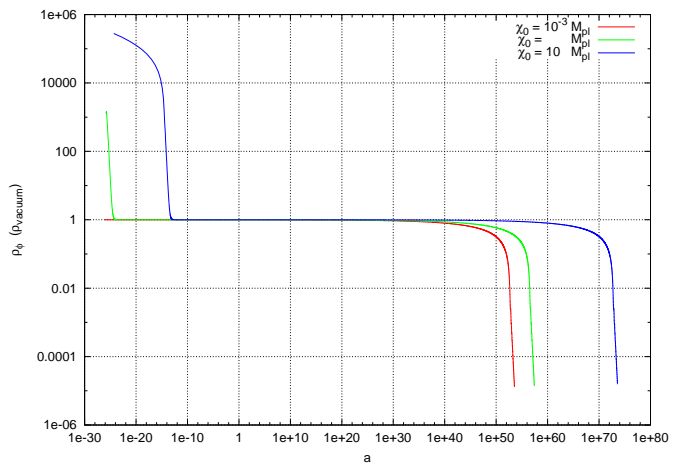}
\caption{The evolution of $\rho_\phi$ for three different initial conditions. Once $\rho_{\chi}$ drops below ${\mu}^2 M^2$, $\phi$ starts rolling towards the critical value $\frac{\mu}{\sqrt{\lambda}}$, and $\phi$ is slowly varying during inflation.}
\label{rho_phi}
\end{figure}

\begin{figure}[h]
\centering
\includegraphics[width=3.5in]{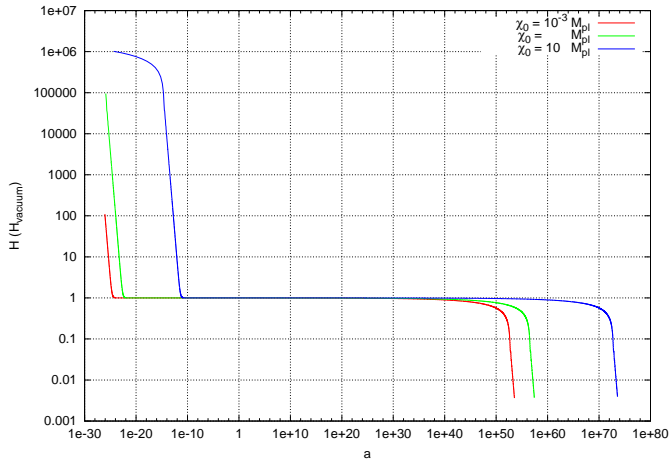}
\caption{The evolution of the Hubble parameter for three different initial conditions. $H(t)$ decreases with the scale factor like $a^{-\frac{3}{2}}$ both during the matter dominated era and the end of inflation, when $\phi$ oscillates like a matter field. During inflation, the vacuum energy density dominates over other sources of energy density, so $H$ is nearly constant.}
\label{hubble}
\end{figure}

Before doing the numerical calculation, we need to specify the relevant parameters. From the constraints on the single $\phi$-field inflation model by Planck + WP + BAO we have $\phi_0 \gtrsim 13 M_{pl}$~\cite{Planck}. So we take $\phi_0 = 13 M_{pl}$ in numerical calculations. Another constraint comes from the under-damping of $\chi$ during the matter-dominated era. If we omit the $\phi$-related terms in the equation of motion of $\chi$ in Eq.~(\ref{eom_chi}), the constraint becomes $H \ll m$. The total energy density of the system decreases towards $\rho_{vacuum}$ before inflation. So we have
\begin{equation}\label{inequality}
\sqrt{\rho_{vacuum}/M_{pl}^2} \ll \sqrt{\rho_{total}/M_{pl}^2} \ll m.
\end{equation}
From this we get $m \gg 10\mu$. In the numerics, we set $\frac{m}{\mu}=10^6$.
  We can now numerically solve the coupled equations ~(\ref{Friedmann}), ~(\ref{eom_phi}) and ~(\ref{eom_chi}). To verify the self-consistency of the solution, we make plots of $\rho_\chi$ and $H$, which should evolve as $a^{-3}$ and $a^{-3/2}$ respectively during the matter-dominated phase. As shown in Fig.~(\ref{rho_chi}) and Fig.~(\ref{hubble}), this behavior is reproduced after thermal equilibrium is reached. Soon after that, $\rho_{\chi}$ drops below the critical values ${\mu}^2 M^2$, which triggers the phase transition. At this moment, the symmetron field $\phi$ field starts rolling down slowly, as shown in Fig.~(\ref{rho_phi}). During inflation the total energy density comes mainly from $\rho_\phi$, in which $\rho_{vacuum}$ is the largest contribution before inflation ends. Therefore $H$ is nearly constant during this time. Then at the end of inflation, $\phi$ quickly drops to $\phi_0$, and $\rho_\phi$ is driven to zero. Finally after inflation, $\phi$ oscillates around $\phi_0$ like a matter field, which makes $H$ and $\rho_\phi$ behaving in the same way as $H$ and $\rho_\chi$, respectively, in the matter-dominated universe. We have quite a large freedom to set the initial value of $\phi$ to any reasonable value close to zero without changing this slow-roll behavior. So without any loss of generality, we set it to be $0.01 \phi_0$ (where $\phi_0 =13 M_{pl}$) in the initial conditions. On the other hand, we consider different initial values of $\chi$, larger than the symmetry-breaking value of $10^{-5}M_{pl}$, in order to verify the robustness of initial conditions. Clearly, they just shift the time of the start and end of inflation, without changing the evolution behavior of the two fields.

\begin{figure}[h]
\centering
\includegraphics[width=3.5in]{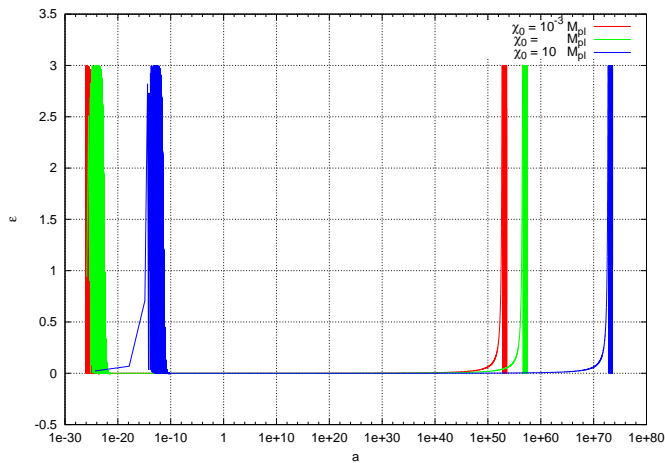}
\caption{The evolution of the slow-roll parameter $\epsilon$ for three different initial conditions. $\epsilon$ is oscillating with an average value of $\frac{3}{2}$ both during the matter dominated era and the end of inflation. During inflation, $\epsilon$ is slowly changing with time.}
\label{epsilon}
\end{figure}

Two important things to check in this context are the number of e-folds and the magnitude of the slow-roll parameter $\epsilon = -\frac{\dot{H}}{H^2}$. $\epsilon$ must be small and slowly varying compared with the cosmic expansion rate. The most recent constraint from Planck is $\epsilon < 0.053$ ~\cite{Planck}. From Fig.~(\ref{epsilon}), we see that $\epsilon$ always satisfies this constraint during inflation in our model. The middle flat part on this plot corresponds to the inflationary phase, so we can see from the change of scale factor that the number of e-folds is far larger than $60$ for any initial conditions.

We therefore see that our inflationary scenario satisfies the usual inflationary requirements without any special initial conditions.

\section{Power spectrum}

As in any model of inflation, much important information about the symmetron inflation is encoded in the power spectrum for primordial perturbations.
Our model, defined by the general action in Eq.~(\ref{action}) and specific matter Lagrangian in Eq.~(\ref{ML}), is just a concrete realization of the multi-field inflation. As derived in Chapter $10$ of \cite{weinberg}, for this kind of potential, the two point correlation function of the invariant quantity ${\cal R}$ in the Bunch-Davies vacuum is
\be
\int d^4 x e^{-i \vec{q}(\vec{x}-\vec{y})}\left\langle {\cal R}(\vec{x},t){\cal R}(\vec{y},t) \right\rangle =(2\pi)^3 \sum_N |{\cal R}_{N_q}|^2 ,
\ee
where $q$ is the comoving wave number of perturbations.
The root-mean-square value of $|{\cal R}_{N_q}|$ is
\begin{equation} \label{Rnq}
\left(\sum{|{\cal R}_{N_q}|^2} \right)^{1/2} = \frac{H^2}{2 M_{pl}(2\pi q)^{3/2}\sqrt{|\dot{H}|}} ,
\end{equation}
where $H$ is the Hubble parameter. This implies that the form of the two-point correlation function at the end of horizon exit will be the same as in the single-field inflation.
In our concrete case, the Hubble parameter $H^2  = \frac{1}{3 M_{pl}^2} \rho$ is given in Eq.~(\ref{Friedmann}).
During inflation, $H$ will be dominated by the potential term in Eq.~(\ref{Friedmann}).
As explained in \cite{weinberg}, the result in Eq.~(\ref{Rnq}) can be further simplified by noticing that the matrix
\be
M^n_l \equiv \gamma^{nm} \frac{\partial^2\bar{V}}{\partial \bar{\varphi}^m \partial \bar{\varphi}^l },
\ee
where $\gamma^{nm}$ is the metric in the field space and $\bar{\varphi}^m = (\bar{\phi},\bar{\chi})$  (bar over the field denotes its unperturbed value), has only one small eigenvalue. In our concrete case
\be
\gamma^{nm} = \begin{pmatrix}
  1     & 0\\
 0 & A^2(\phi)
\end{pmatrix},
\ee

\be
\bar{V}=V(\phi) + \frac{1}{2}A^4m^2\chi^2 ,
\ee

from where we get

%\bea
%&&  \frac{\partial^2 \bar{V}}{\partial \phi^2} = \frac{\partial^2 \bar{V}}{\partial\phi^2} +2m^2\left[ A^3 \frac{\partial^2 A}{\partial\phi^2} +3A^2 \left( \frac{\partial A}{\partial \phi} \right)^2 \right]\chi^2 ,  \\
%&&  \frac{\partial^2 \bar{V}}{\partial \chi \partial \phi} = 4m^2A^3 \left( \frac{\partial A}{\partial \phi} \right) \chi , \\
%&&  \frac{\partial^2 \bar{V}}{\partial \chi^2} = mA^4 .
%\eea
\begin{eqnarray}
&&  \frac{\partial^2 \bar{V}}{\partial \phi^2} = \frac{\partial^2 \bar{V}}{\partial\phi^2} +2m^2\left[ A^3 \frac{\partial^2 A}{\partial\phi^2} +3A^2 \left( \frac{\partial A}{\partial \phi} \right)^2 \right]\chi^2 ,  \cr
&&  \frac{\partial^2 \bar{V}}{\partial \chi \partial \phi} = 4m^2A^3 \left( \frac{\partial A}{\partial \phi} \right) \chi , \cr
&&  \frac{\partial^2 \bar{V}}{\partial \chi^2} = mA^4 .
\end{eqnarray}

Thus, the matrix $M^n_l$ is

\be
M^n_l = \begin{pmatrix}
  \frac{\partial^2 \bar{V}}{\partial \phi^2}     & \frac{\partial^2 \bar{V}}{\partial \chi \partial \phi}\\
 A^2 \frac{\partial^2 \bar{V}}{\partial \chi \partial \phi} & A^2\frac{\partial^2 \bar{V}}{\partial \chi^2}
\end{pmatrix}.
\ee
We saw that during inflation the field $\chi$ is damped, so we can set $\chi \approx 0$, in which case we get
\be
M^n_l \approx \begin{pmatrix}
  -\mu^2+3\lambda \phi^2     & 0\\
 0 & m^2A^6
\end{pmatrix}.
\ee
Using the inequality $\phi \leq \mu/\sqrt{\lambda}$, we can limit the first eigenvalue of $M^n_l$
\be
-\mu^2+3\lambda \phi^2 \leq 2\mu^2 ,
\ee
while the second one is
\be
m^2A^2 = m^2 \left(1+\frac{\phi^2}{2M^2}  \right) \leq  m^2 \left(1+\frac{\mu^2}{2\lambda M^2}  \right) = \frac{3}{2}m^2 .
\ee
A generic value that we used in our numerical calculations $m^2/\mu^2 =10^{12}$ indicates that the first eigenvalue is much smaller than the second one.
Then the unperturbed field rolls only along one direction (of the small eigenvalue) and the only significant perturbations lie in that direction. The results then reduce to the single field slow roll inflation. The amplitude of the power spectrum is
\be
{\cal R}_q^0\propto q^{-3/2 -2\epsilon -\delta} ,
\ee
where $\epsilon \equiv -\dot{H}/H^2$ and $\delta \equiv \ddot{H}/{2H\dot{H}}$ are the slow roll parameters. In this approximation the slope is given by
\be
n_S(q) = 1-4\epsilon - 2 \delta,
\ee
which, in the single-field($\phi$) slow-roll approximation, can be written explicitly in terms of our model parameters as
\be
n_S = 1-\frac18\left(\frac{M_{pl}}{\phi_0} \right)^2 \frac{x^2_N}{(x^2_N-1)^2} + 8\left(\frac{M_{pl}}{\phi_0} \right)^2 \frac{1}{x^2_N-1},
\ee
where $x_N=\frac{\phi_N}{\phi_0}$ with $\phi_N$ being the field value at the number of e-folds N. If we take $x_N=0$ and $n_S\approx 0.96$, we can get the approximate value of $\phi_0=\mu/\sqrt{\lambda}\approx 14M_{pl}$.

Since the only role of multiple fields here is to set initial conditions, the amplitude of the tensor perturbations at the horizon exit can be written in the usual form for the single-field inflation,
\be
{\cal D}_q^0 = i \frac{H}{M_{pl} (2\pi q)^{3/2}} ,
\ee
and the slope parameter is
\be
n_T(q) = -2\epsilon ,
\ee
just as in the single field case.

\begin{figure}[h]
\centering
\includegraphics[width=3.5in]{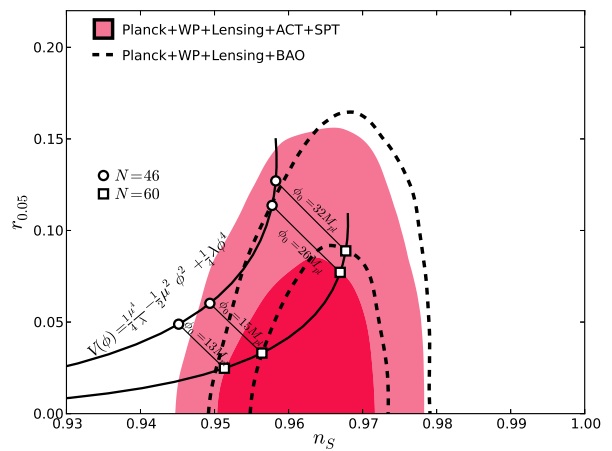}
\caption{The constraint on the parameter $\phi_0=\mu/\sqrt{\lambda}$ from observations of $r_{0.05}$ and $n_S$. The darker shadow and the inner dashed lines represent the $1\sigma$ constraint area, while the lighter shadow and the outer dashed lines represent the $2\sigma$ area. The trends of both the $N=46$ and $N=60$ lines are shown, as well as the $1\sigma$ constraints on $\phi_0$ in our model.}
\label{planck}
\end{figure}

Therefore, treating our theory as a single-field($\phi$) inflation model, we can constrain our model parameters with the latest Planck data and other cosmological observations, as shown in Fig.~(\ref{planck}). Here we used the second-order slow-roll approximation \cite{2-ed_slow_roll}, which gives the ratio of scalar to tensor perturbations and the scalar spectral index as
\begin{align}
r(N)\equiv & \frac{|{\cal D}_q^0|^2}{|{\cal R}_q^0|^2} = 16\epsilon(\phi_N)\left(1-C(\sigma(\phi_N)+2\epsilon(\phi_N)) \right) \\
n_S(N)=& 1+\sigma(\phi_N)-(5-3C)\epsilon^2(\phi_N) \nonumber\\
     &-\frac14(3-5C)\sigma(\phi_N)\epsilon(\phi_N)-\frac12(3-C)\delta(\phi_N),
\end{align}
where $C=4(ln2+\gamma)-5\approx 0.0814514$, with $\gamma\approx0.577$ being Euler's constant. Here N is the number of e-folds and $\sigma(\phi_N) = -2\delta-4\epsilon$.

The above parameters are taken at the pivot scale $k_*=0.002 Mpc^{-1}$ by default. In order to reduce the degeneracy between spectral indices and amplitudes, we chose another pivot scale $k_*=0.05 Mpc^{-1}$. The $r(N)$ parameters at these two scales are related by
\begin{equation}
r_{0.05}(N) = r(N) \left( \frac{0.05}{0.002}\right)^{1-n_S(N)-r(N)/8}.
\end{equation}

It turns out that the $1\sigma$ constraint on $\phi_0=\mu/\sqrt{\lambda}$ from the combination of Planck, WP, Lensing, ACT and SPT data is $13 M_{pl} \le \phi_0 \le 26 M_{pl}$, while that from the combination of Planck, WP, Lensing and BAO data is $15M_{pl} \le \phi_0 \le 32M_{pl}$. Our result is consistent with that given by the Planck group \cite{Planck}.

\section{Conclusions}
In this paper we defined an inflationary scenario in the context of symmetron models with the scalar field whose potential depends on the energy density of its environment.  In our model inflation starts naturally when the energy density of the universe drops below some critical value.
We introduced an auxiliary field which is coupled to the inflaton field and describes the matter content of the universe before the onset of inflation. The dynamics of the model is such that this auxiliary field quickly drops to its minimum after the onset of inflation and has no significant influence in the inflationary phase. Thus, during inflation we have the standard single field slow roll behavior. We verified that our model is consistent with usual requirements and cosmological observations, including the newest Planck data.

We note that this ``symmetron inflation" shares some features with  ``new inflation” in \cite{Albrecht:1982wi,Linde:1981mu}. ``New inflation" is a second order phase transition
where temperature dependent corrections change the shape of the potential. While ``symmetron inflation" is also a second order phase transition, the model is based on modified gravity and the role of temperature is played by the energy density of the universe. Thus inflation can start even in a matter dominated era.
Like in ``new inflation", the most generic initial conditions near the cosmological singularity are anisotropic and inhomogeneous (the scalar field is generically non-zero, large and inhomogeneous). Thus, if isotropy, homogeneity and the thermal equilibrium have been reached sufficiently fast, before the
space-time curvature has fallen below the scale of inflation, then no other special initial conditions are needed for the onset of inflation.

As in any other scalar field model, potential danger comes from quantum corrections.  Scalar fields are always sensitive to corrections due to interactions with other fields, with the standard model hierarchy problem as the most notable example. To compute exact form of corrections in our model would be  difficult since the fields $\chi$ and  $\phi$ interact not only through their potentials but also through the derivatives. However we do expect  the corrections to become negligible once inflation starts since the field $\chi$ would be practically frozen at its minimum at $\chi =0$.

\begin{acknowledgments}
The authors thank Justin Khoury for very useful discussions. This research is supported in part by the National Science
Foundation under the grant NSF-PHY-1066278.
\end{acknowledgments}

\appendix*
\section{Action in the Jordan frame}

In order to put the total action (\ref{action}) in the Jordan frame, we just need to rewrite Einstein-frame Ricci scalar $R$ in terms of the Jordan-frame counterpart $\tilde{R}$. For any conformal transformation $\tilde{g}_{\mu\nu} = F(x)g_{\mu\nu}$, it is straightforward to get
\begin{align}
\tilde{R}= &\frac1F \left(R+ 3g^{\mu\kappa}\frac{F_{,\mu,\kappa}}{F}- \frac32 g^{\mu\kappa}\frac{F_{,\mu}F_{,\kappa}}{F^2}\right.\nonumber\\
         &+\left.\frac{F_{,\rho}}{F}\left[3\frac{\partial g^{\lambda\rho}}{\partial x^\lambda}+ \frac32 g^{\eta\rho}g^{\lambda\omega}\frac{\partial g_{\lambda\omega}}{\partial x^\eta} \right] \right).
\label{tildeR}
\end{align}

In our case, $F=A^2(\phi)$, we can rewrite each term in the original action as follows,
\begin{align}
\frac12\sqrt{-g}M^2_{pl}R = &\frac12\sqrt{-\tilde{g}} M^2_{pl}\left(\frac{\tilde{R}}{A^2}-6\tilde{g}^{\mu\kappa}\frac{A_{,\mu}A_{,\kappa}}{A^3}+12\tilde{g}^{\lambda\rho}\right.\nonumber\\
&\left.\times\frac{A_{,\rho}A_{,\lambda}}{A^4}-6\frac{\partial \tilde{g}^{\lambda\rho}}{\partial x^{\lambda}}\frac{A_{,\rho}}{A^3}-3\tilde{g}^{\eta\rho}\tilde{g}^{\lambda\omega}\frac{\partial \tilde{g}_{\lambda\omega}}{\partial x^{\eta}}\frac{A_{,\rho}}{A^3} \right) \\
-\frac12\sqrt{-g}g^{\mu\nu}\partial_\mu&\phi\partial_\nu\phi = -\frac12\sqrt{-\tilde{g}}\tilde{g}^{\mu\nu}\frac1{A^2}\partial_\mu\phi\partial_\nu\phi \\
-\sqrt{-g}V(\phi) =& -\sqrt{-\tilde{g}}\frac{V(\phi)}{A^4}.
\end{align}

Finally, the whole action can be written in the Jordan frame in the following form,
\begin{align}
S=&\int d^4x \sqrt{-\tilde{g}}\left(\frac1{2A^2}M^2_{pl}\tilde{R}-\frac1{2A^2}\tilde{g}^{\mu\nu}\partial_\mu\phi\partial_\nu\phi -\frac{V(\phi)}{A^4}\right.\nonumber\\
&-3M^2_{pl}\frac{A_{,\mu,\nu}}{A^3}\tilde{g}^{\mu\nu}+6M^2_{pl}\frac{A_{,\mu}A_{,\nu}}{A^4}\tilde{g}^{\mu\nu}-3M^2_{pl}\frac{A_{,\nu}}{A^3}\frac{\partial \tilde{g}^{\mu\nu}}{\partial x^\mu} \nonumber\\
&\left.-\frac32M^2_{pl}\frac{A_{,\nu}}{A^3}\tilde{g}^{\mu\nu}\tilde{g}^{\lambda\omega}\frac{\partial \tilde{g}_{\lambda\omega}}{\partial x^\mu} + {\mathcal L}_m(\chi,\tilde{g}_{\mu\nu}) \right).
\end{align}

\end{document}